\documentclass[12pt]{iopart}
\usepackage{graphicx}

\begin{document}

\title{Magnetism study on a triangular lattice antiferromagnet Cu$_2$(OH)$_3$Br}

\author{Z Y Zhao$^1$, H L Che$^2$, R Chen$^3$, J F Wang$^3$, X F Sun$^{2,4,5}$ and Z Z He$^1$}

\address{$^1$ State Key Laboratory of Structural Chemistry, Fujian Institute of Research on the Structure of Matter, Chinese Academy of Sciences, Fuzhou, Fujian 350002, People's Republic of China}

\address{$^2$ Department of Physics, Hefei National Laboratory for Physical Sciences at Microscale, and Key Laboratory of Strongly-Coupled Quantum Matter Physics (CAS), University of Science and Technology of China, Hefei, Anhui 230026, People's Republic of China}

\address{$^3$ Wuhan National High Magnetic Field Center, Huazhong University of Science and Technology, Wuhan, Hubei 430074, People¡¯s Republic of China}

\address{$^4$ Institute of Physical Science and Information Technology, Anhui University, Hefei, Anhui 230601, People's Republic of China}

\address{$^5$ Collaborative Innovation Center of Advanced Microstructures, Nanjing, Jiangsu 210093, People's Republic of China}

\ead{hezz@fjirsm.ac.cn}
\ead{xfsun@ustc.edu.cn}


\begin{abstract}
Magnetism of Cu$_2$(OH)$_3$Br single crystals based on a triangular lattice is studied by means of magnetic susceptibility, pulsed-field magnetization, and specific heat measurements. There are two inequivalent Cu$^{2+}$ sites in an asymmetric unit. Both Cu$^{2+}$ sublattices undergo a long-range antiferromagnetic (AFM) order at $T\rm_N$ = 9.3 K. Upon cooling, an anisotropy crossover from Heisenberg to $XY$ behavior is observed below 7.5 K from the anisotropic magnetic susceptibility. The magnetic field applied within the $XY$ plane induces a spin-flop transition of Cu$^{2+}$ ions between 4.9 T and 5.3 T. With further increasing fields, the magnetic moment is gradually increased but is only about half of the saturation of a Cu$^{2+}$ ion even in 30 T. The individual reorientation of the inequivalent Cu$^{2+}$ spins under field is proposed to account for the magnetization behavior. The observed spin-flop transition is likely related to one Cu site, and the AFM coupling among the rest Cu spins is so strong that the 30-T field cannot overcome the anisotropy. The temperature dependence of the magnetic specific heat, which is well described by a sum of two gapped AFM contributions, is a further support for the proposed scenario.

\end{abstract}

\vspace{2pc}
\submitto{\JPCM}

%

\section{Introduction}

Geometrical frustration embedded in quantum antiferromagnets can introduce numerous unconventional magnetism such as spin ice, spin liquid, and magnetization plateau \cite{1,2,3,4,5,6,7,8}. Thereinto, spin liquid is the most attractive subject in modern condensed matter physics \cite{9,10,11,12,13,14,15,16}. The first reported candidate is herbersmithite ZnCu$_3$(OH)$_6$Cl$_2$ \cite{17}, and its parent compound clinoatacamite Cu$_2$(OH)$_3$Cl is also a frustrated antiferromagnet \cite{18}. Cu$_2$(OH)$_3$Cl belongs to $M_2$(OH)$_3X$ ($M$ = transition metal; $X$ = Cl, Br, I) family, which has four frustrated polymorphs named as atacamite, botallackite, clinoatacamite, and paratacamite. When replacing different transition metals or halogens, diverse spin fluctuations and exchange interactions can compete with the structure-related frustration effect and result in variously novel ground states. $M_2$(OH)$_3X$ is therefore a good material playground to study the geometrical frustration and the associated quantum magnetism. In addition, strong magnetic-dielectric-lattice coupling was also observed in $M_2$(OH)$_3X$ signaling the multiferroic behavior and a potential application in magnetic storage \cite{19}.

Investigation of anisotropic magnetism on single crystals is essential to explore physics of $M_2$(OH)$_3X$ family. However, early studies on $M_2$(OH)$_3X$ were mostly performed on polycrystals or microcrystals. Though there exist natural $M_2$(OH)$_3X$ crystals, the purity and size were not sufficient for the exploration of the intrinsic anisotropic magnetism. The lack of large synthetic crystals impedes the deep understanding of the physics from two aspects. First, the magnetism is usually different in powder and crystals. In atacamite Cu$_2$(OH)$_3$Cl, the synthetic powder showed a spin-glass behavior while the mineral crystal underwent a antiferromagnetic (AFM) transition at $T\rm_N$ = 9 K \cite{20,21}. In paratacamite Co$_2$(OH)$_3$Cl polycrystalline sample, besides the kagom\'{e}-ice transition at $T\rm_C$ = 10.5 K as observed in single crystals, a glasslike freezing was also found below 3 K which was likely resulted from the micro Co deficiency \cite{22}. Even for powder, paratacamite Fe$_2$(OH)$_3$Cl synthesized by different groups showed obviously different transition temperatures \cite{23,24}. Second, the behavior observed in powder sample is an average of the anisotropic properties which is not able to offer accurate physical information. For example, two successive field-induced transitions and a possible magnetization plateau were detected in paratacamite Co$_2$(OH)$_3$Br \cite{25}. A completely flat plateau was suspected to show in the anisotropic magnetization performed on a single crystal. For paratacamite Fe$_2$(OH)$_3$Cl, neutron diffraction gave four possible magnetic structures, and it was pointed out that only single-crystal experiment can determine the definite spin configuration \cite{24}.

\begin{figure}[h]
\centering
\includegraphics[height=6.5cm]{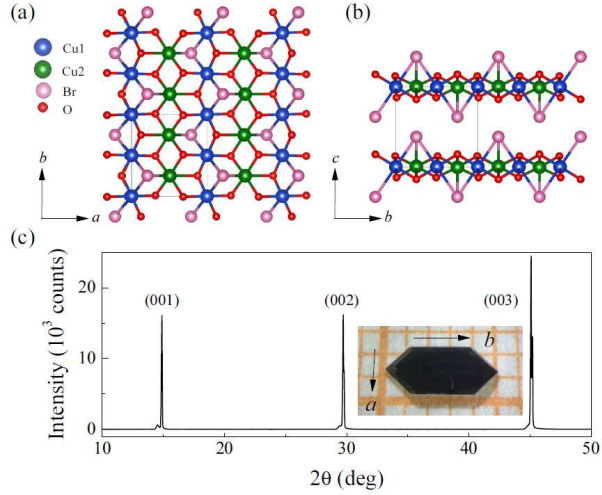}
\caption{(a-b) Crystal structures projected in the $ab$ plane and viewed along the $a$ axis, respectively. Hydrogen atoms are omitted for clarity. (c) X-ray diffraction on the (00$l$) facet. Inset is a photograph of one piece of botallackite Cu$_2$(OH)$_3$Br single crystal. The $a$ and $b$ axes are labeled as arrows.}
\end{figure}

In this work, we successfully synthesized large and high-quality botallackite Cu$_2$(OH)$_3$Br single crystals for the first time. Among all polymorphs, botallackite is the rarest in nature and little study has been thus carried out. Different from the other three polymorphs in which the magnetic ions constitute a pyrochlore-like network, botallackite has a two dimensional (2D) triangular lattice. Figures 1(a) and 1(b) display the crystal structure of botallackite Cu$_2$(OH)$_3$Br. There are two peculiar Cu$^{2+}$ positions in an asymmetric unit. Cu1 is octahedrally coordinated by four hydroxyls and two Br atoms whereas Cu2 is surrounded with five hydroxyls and one Br atom, and both copper octahedra are strongly distorted. The Cu-O bonds constitute a 2D sheet parallel to $ab$ plane, and the Br atoms are located on both sides of each sheet. The 2D sheet can be considered as a triangular lattice composed of edge-shared Cu1 and Cu2 uniform chains along the $b$ axis which are further connected alternately by sharing two edges along the $a$ axis. These sheets are stacked along the $c$ axis through O-H$\cdot\cdot\cdot$Br bonds so as to develop a three-dimensional network. The resultant weak interlayer interaction accounts for the good two dimensionality, and botallackite Cu$_2$(OH)$_3$Br is thus regarded as an $S$ = 1/2 spatial anisotropic triangular lattice compound. As far as we know, only one work performed on botallackite Cu$_2$(OH)$_3$Br polycrystalline sample was reported \cite{26}. A broad peak was observed around 10 K in the magnetic susceptibility, which was ascribed to the long-range AFM transition of Cu$^{2+}$ ions.

In this manuscript, the magnetism of botallackite Cu$_2$(OH)$_3$Br single crystals is studied. A slope change of the magnetic susceptibility is observed at $T\rm_N$ = 9.3 K which is associated with the long-range AFM order as confirmed by the $\lambda$ peak in specific heat, while the broad peak around 10 K is resulted from the development of the short-range AFM correlation. With lowering temperature, a temperature-induced anisotropy crossover from Heisenberg to $XY$ behavior is observed below 7.5 K. When the magnetic field is applied within the $XY$ plane, a spin-flop transition occurs at low fields, and the magnetic moment is only about half of the saturation of a Cu$^{2+}$ ion even in 30 T. Considering the distinct local environments of Cu1 and Cu2, individual field-induced spin reorientation is proposed to account for the magnetization behavior. The low-field spin-flop transition is likely related to one Cu site, and the anisotropy energy of the other Cu site is too strong to be overcome by the 30-T field. Such scenario can be well explained by the temperature dependence of the magnetic specific heat composed of two gapped AFM contributions with different gaps.

\section{Methods}

Cu$_2$(OH)$_3$Br single crystals were grown using a conventional hydrothermal method. Stoichiometric CuBr$_2$ and Cu(NO$_3$)$_2$$\cdot$3H$_2$O were dissolved in 2 mL deionized water, and then sealed in an autoclave equipped with a 28 mL Teflon liner. The autoclaves were heated at 230$^\circ$C for 4 days under autogenous pressure and then cooled to room temperature at a rate of 1.5 $^\circ$C/h for 6 days. The as-grown single crystals are dark green with elongated hexagon shape, as shown in the inset to Fig. 1(c). The typical size of the crystals is 5 mm $\times$ 2 mm $\times$ 0.5 mm.

Single crystal x-ray diffraction was collected on a Rigaku Mercury CCD diffractometer equipped with a graphite-monochromated Mo-K$\alpha$ radiation ($\lambda$ = 0.71 {\AA}) at room temperature. The refined lattice parameters $a$ = 5.6613(11) {\AA}, $b$ = 6.1596(7) {\AA}, $c$ = 6.0829(9) {\AA}, and $\beta$ = 93.569(15)$^\circ$ are consistent with previously reported \cite{26}. The largest facet of the as-grown single crystal was checked to be parallel to the $ab$ plane as seen from the x-ray diffraction on the (00$l$) facet in Fig. 1(c), and the length direction was further confirmed to be along the $b$ axis. In this work, the direction perpendicular to the $ab$ plane is defined as $c^*$ axis.

Magnetic susceptibility and magnetization were measured using a SQUID-VSM (Quantum Design) between 2 and 300 K up to 7 T. Specific heat was measured by the relaxation method between 2-150 K using a PPMS (Quantum Design). Pulsed-field magnetization was performed at 2 K up to 30 T on a self-built platform in Wuhan National High Magnetic Field Center (China).

\section{Results and Discussion}

\subsection{Magnetic properties}

\begin{figure}[h]
\centering
\includegraphics[height=6cm]{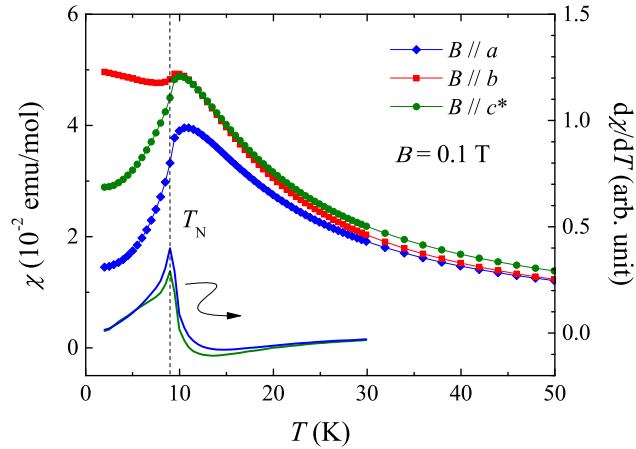}
\caption{Temperature dependencies of the magnetic susceptibility measured in 0.1 T along the three directions. The differentials of the magnetic susceptibility along the $a$ and $c^*$ axes are also plotted.}
\end{figure}

Temperature dependencies of the magnetic susceptibility $\chi(T)$ measured along the three directions in $B$ = 0.1 T are shown in Fig. 2. No difference is observed in the field-cooling (FC) and zero-field-cooling (ZFC) measurements. Above 100 K, $\chi(T)$ follows well the Curie-Weiss law $\chi$ = $\chi_0$ + $C/(T - \theta\rm_{CW}$), where $\chi_0$ is a temperature-independent term. The Curie-Weiss temperatures for three directions are $\theta_{{\rm{CW}},a}$ = -21.7(7) K, $\theta_{{\rm{CW}},b}$ = -15.5(4) K, $\theta_{{\rm{CW}},c^*}$ = -19(1) K, and the effective moments are deduced to be $\mu_{{\rm{eff}},a}$ = 1.80 $\mu\rm_B$/Cu$^{2+}$, $\mu_{{\rm{eff}},b}$ = 1.73 $\mu\rm_B$/Cu$^{2+}$, $\mu_{{\rm{eff}},c^*}$ = 1.89 $\mu\rm_B$/Cu$^{2+}$, respectively. The negative $\theta\rm_{CW}$ suggests a dominant AFM interaction among Cu$^{2+}$ spins. Upon cooling, $\chi(T)$ is gradually increased and exhibits a broad peak around 10 K. This is a common feature in the low-dimensional antiferromagnets and indicates the development of short-range AFM correlation. With further lowering temperature, $\chi_a$ and $\chi_{c^*}$ show a sudden drop at about 9 K, which is associated with the AFM order of Cu$^{2+}$ spins on account of the inevitable interlayer interaction. The transition temperature is defined as the peak position of the differentials in Fig. 2. Apparently, the broad peak observed around 10 K in the previous polycrystalline study is related to the short-range AFM correlation rather than the AFM order \cite{26}. The absence of the anomaly across $T\rm_N$ in $\chi(T)$ in that work might be due to a different sample quality in different growth condition, which further highlights the importance of single crystals in understanding the physics of the quantum antiferromagnets.

At lower temperatures, a minimum is observed in $\chi_b$ at 7.5 K. The presence of the minimum is a remarkable feature for the exchange-anisotropy crossover from Heisenberg to $XY$ behavior in two-dimensional antiferromagnets \cite{6,27,28,29,30,31}. When the temperature is decreased, the $XY$ anisotropy becomes significant and a large amount of spins are confined in the plane. As a result, the in-plane $\chi(T)$ decreases faster, while the perpendicular component is reduced with canted ferromagnetic moment along the field direction. Consequently, a minimum finally occurs in the out-of-plane component, and the minimum position is usually defined to be the crossover temperature below which the AFM correlation becomes irrelevant in the out-of-plane component. Since the minimum appears in $\chi_b$, the magnetic $XY$ plane is therefore the $ac^*$ plane which is not identical to the triangular layer from the crystal structure determined at room temperature. X-ray diffraction performed below 7.5 K is demanded to investigate the crystal structure and explore the origin of the exotic $XY$ anisotropy.

\begin{figure}[h]
\centering
\includegraphics[height=12cm]{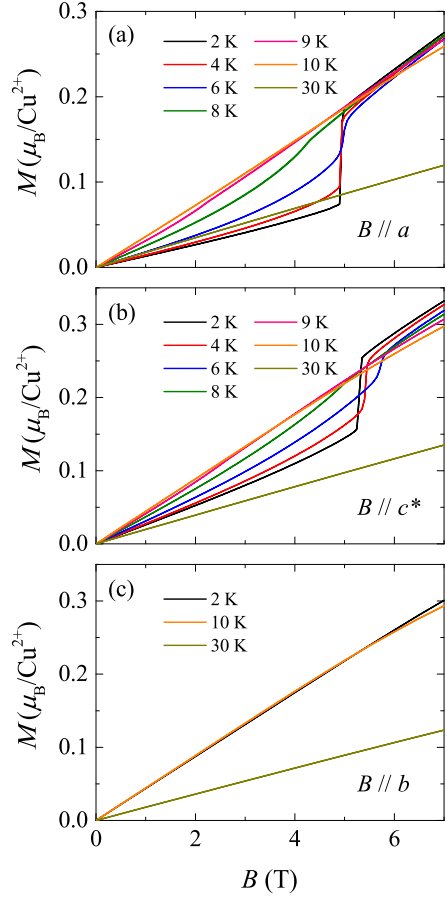}
\caption{Magnetic-field dependencies of the magnetization along three directions at different temperatures.}
\end{figure}

Figure 3 shows the magnetization $M(B)$ measured in static fields up to 7 T along the three directions. At 2 K, both $M_a$ and $M_{c^*}$ exhibit a step-like enhancement with increasing magnetic field. Upon warming, the critical field for $M_a$ is almost unchanged ($\sim$ 4.9 T) below 6 K, while it is increased from 5.3 T at 2 K to 5.6 T at 6 K for $M_{c^*}$. Since the applied field is within the $XY$ plane, the step-like enhancements observed in $M_a$ and $M_{c^*}$ are likely resulted from the spin flop of Cu$^{2+}$ spins. However, the different tendencies of the critical fields deserve further investigations to explore the difference between $M_a$ and $M_{c^*}$. In contrast, $M_b$ shows a linear increase with the magnetic field.

\begin{figure}[h]
\centering
\includegraphics[height=8cm]{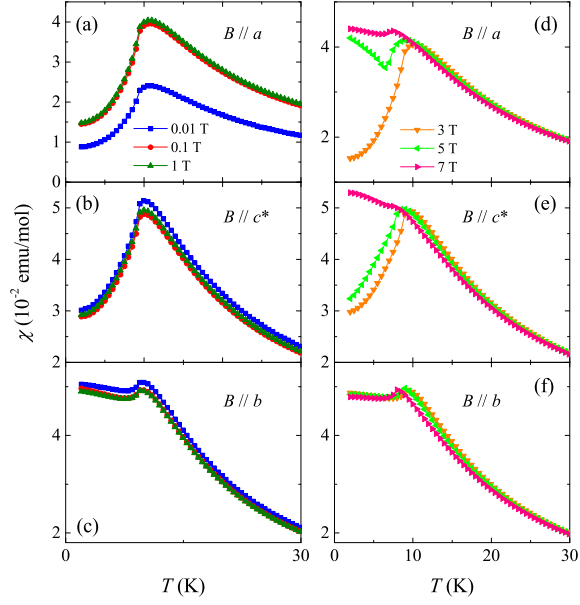}
\caption{Anisotropic magnetic susceptibilities measured in various magnetic fields below 1 T (a-c) and above 3 T (d-f) along the three directions.}
\end{figure}

The anisotropic magnetic susceptibilities measured up to 7 T are displayed in Fig. 4. At low fields, both $\chi_a$ and $\chi_{c^*}$ are quickly decreased at low temperatures due to the $XY$ anisotropy. When the field is higher than the spin-flop transition field, the low-temperature magnetic susceptibility is increased resulting in a minimum which corresponds to the transition from paramagnetic state to the spin-flop state. On the other hand, the behavior of $\chi_b$ is almost unchanged in the magnetic fields suggesting a robust $XY$ anisotropy.

\begin{figure}[h]
\centering
\includegraphics[height=5.5cm]{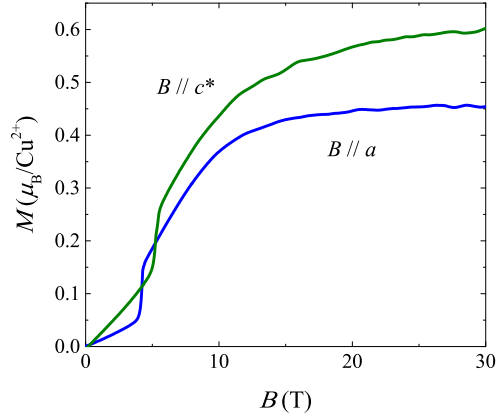}
\caption{Magnetization measured in pulsed fields up to 30 T at 2 K with field applied in the $XY$ plane.}
\end{figure}

From Fig. 3 it is clearly seen that $M_{c^*}$ is largest as compared with the other two components, while it is only about 0.33 $\mu\rm_B$/Cu$^{2+}$ in 7 T and much smaller than the saturated moment of a Cu$^{2+}$ ion. To explore the possible magnetic phase transitions in higher fields, pulsed-field magnetization is performed up to 30 T at 2 K. As shown in Fig. 5, except for the spin-flop transition at lower fields, there is no extra transition found in higher fields when the magnetic field is applied in the $XY$ plane. In spite of the slow increase above 10 T, there should not be magnetization plateau in Cu$_2$(OH)$_3$Br. In triangular lattices, the 1/3 magnetization plateau is stabilized by the quantum fluctuations, and an upward concave curvature is usually appeared before entering into the plateau phase. The rounding of the magnetization in Cu$_2$(OH)$_3$Br seems more like the spin polarization process. It should be mentioned that the magnetization is a bit sample dependent, that is, the enhancement of the magnetization across the spin-flop transition for the crystals grown from different batches is somewhat different, but the magnitude at high fields is always close or smaller than 0.5 $\mu\rm_B$. Such difference has no significant influence on discussing the ground state in the following section.

\subsection{Specific heat}

\begin{figure}[h]
\centering
\includegraphics[height=12cm]{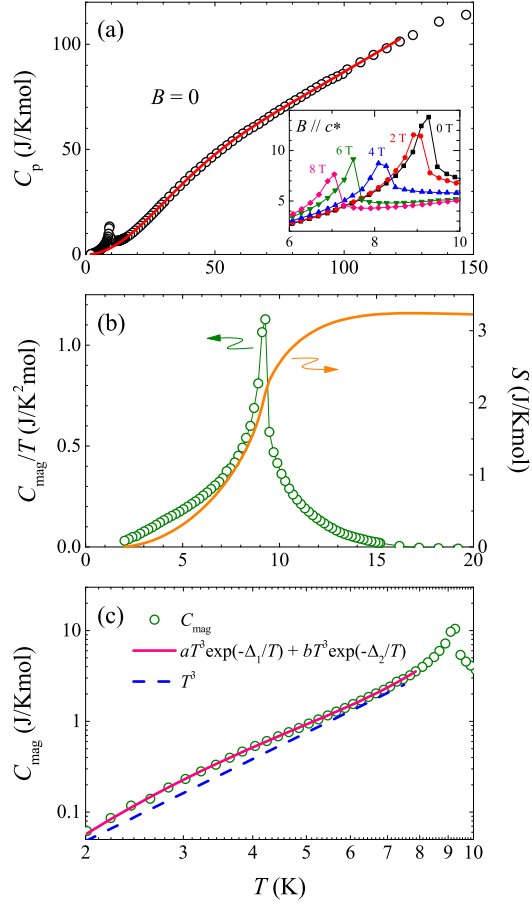}
\caption{Temperature dependence of the specific heat. (a) Total specific heat in zero field. Inset: specific heat in different magnetic fields parallel to the $c^*$ direction. (b) Magnetic specific heat divided by temperature and the entropy change in zero field. (c) Magnetic specific heat in a log-log plot. The solid line is a fit including two gapped AFM contributions. The dashed line is the $T^3$ dependence.}
\end{figure}

Figure 6(a) displays the specific heat $C\rm_p$ measured in zero field. A $\lambda$ peak is observed at $T\rm_N$ = 9.3 K, which is consistent with the magnetic susceptibility and associated with the AFM order. When the field is applied within the $XY$ plane ($B \parallel c^*$ for example), the transition temperature is gradually decreased with reduced magnitude, as shown in the inset to Fig. 6(a). The lattice contribution $C\rm_L$ can be described according to the Thirring approximation \cite{33}
\begin{equation}
C\rm_L = 3\textit{sR}(1 + \sum_{\textit{n}=1}\textit{B}_\textit{n}\textit{u}^\textit{-n}),
\end{equation}
where $u$ = [($T/T\rm_b)^2$ + 1] and $T\rm_b$ = $\theta\rm_D$/2$\pi$, $\theta\rm_D$ is the Debye temperature, $s$ is the number of atoms per molecule, $R$ is the gas constant, and $B_n$ are free parameters. As seen in Fig. 6(a), the experimental curve is well reproduced between 16 and 120 K with the fitting parameters $\theta\rm_D$ = 565(3) K, $B_1$ = -2.43(1), $B_2$ = 3.12(2), $B_3$ = -1.70(1). The magnetic contribution $C\rm_{mag}$ can then be extracted from $C\rm_p$ by subtracting $C\rm_L$. The $C\rm_{mag}$/$T$ and the entropy change as a function of temperature are present in Fig. 6(b). The entropy recovery at 20 K is about 3.2 J/Kmol, a little smaller than the expected $R$ln2 for $S$ = 1/2 systems.

In view of the AFM ground state and no detectable FC/ZFC splitting in $\chi(T)$, Cu1 and Cu2 spins should order antiferromagnetically in different manners. Since the local environments of Cu1 and Cu2 are distinct, the magnetic field may induce the spin-flop transition step by step due to the different anisotropic energies. The step-like enhancement observed in $M(B)$ at low fields is likely related to one Cu site, which is gradually polarized in high fields. The fact that the in-plane magnetic moment is about half of the saturation moment of a Cu$^{2+}$ ion even if the magnetic field is applied as high as 30 T indicates that the anisotropy energy of the other Cu site is too strong to overcome and the rest Cu ions are still in a AFM arrangement.

The conjecture of the separate spin-flop transitions for Cu1 and Cu2 are also supported by the magnetic specific heat. As plotted in Fig. 6(c), $C\rm_{mag}$ deviates from the gapless $T^3$ dependence below $T\rm_N$. Also a single gapped AFM model $C_{\rm{mag}} \sim T^3$exp$(-\Delta/T)$ gives a meaningless energy gap (not shown). Based on the above discussion, $C\rm_{mag}$ is found to be best reproduced by the sum of two gapped AFM contributions with different energy gaps as below \cite{42}
\begin{equation}
C{\rm_{mag}} = aT^3{\rm{exp}}(-\Delta{\rm_1}/T) + bT^3{\rm{exp}}(-\Delta{\rm_2}/T).
\end{equation}
As the solid line shown, the fit yields $\Delta_1$ = 4.7(2) K and $\Delta_2$ = 38(4) K. The yielded $\Delta_2$ is much larger than $\Delta_1$, which means a stronger anisotropy energy and a much higher critical field to rotate the rest Cu spins.

\section{Conclusions}

Single crystals of the triangular lattice antiferromagnet Cu$_2$(OH)$_3$Br are successfully synthesized and the magnetic properties are characterized by means of magnetic susceptibility, pulsed-field magnetization, and specific heat. Inequivalent Cu$^{2+}$ ions are ordered antiferromagnetically at $T\rm_N$ = 9.3 K concurrently. When lowering temperature, an anisotropy crossover from Heisenberg to $XY$ type is observed at 7.5 K, demonstrating that the spins are antialigned in the $ac^*$ plane. When the field is applied within the $XY$ plane, a spin-flop transition occurs at lower fields. The magnetic moment in 30 T is about half of the saturation of a Cu$^{2+}$ ion. A scenario that the inequivalent Cu$^{2+}$ spins individually reorient under magnetic field is proposed to account for the magnetization behavior. A model including two gapped AFM contributions describes well the temperature dependence of the magnetic specific heat. The exotic magnetism in Cu$_2$(OH)$_3$Br is possibly resulted from the delicate balance between the geometrical frustration, spin fluctuations, and complicated exchange interactions.

\section*{Acknowledgements}

This work was supported by the National Natural Science Foundation of China (NSFC) (Grant Nos. U1832166, 51702320, 21573235, and U1632159), the Chinese Academy of Sciences (CAS) under Grant No. KJZD-EW-M05, and the Opening Project of Wuhan National High Magnetic Field Center (Grant No. 2015KF08). XFS acknowledge support from NSFC (Grant Nos. U1832209 and 11874336) and the National Basic Research Program of China (Grant Nos. 2015CB921201 and 2016YFA0300103). JFW acknowledge support from NSFC (Grant No. 11574098).

{}


\section*{References}

\begin{thebibliography}{}

\bibitem{1}
Lacroix C, Mendels P and Mila F 2011 \textit{Introduction to Frustrated Magnetism: Materials, Experiments, Theory} (Springer).

\bibitem{2}
Moessner R and Ramirez A P 2006 \textit{Phys. Today} \textbf{59} 24.

\bibitem{3}
Gardner J S, Gingras M J P and Greedan J E 2010 \textit{Rev. Mod. Phys.} \textbf{82} 53.

\bibitem{4}
Gingras M J P and McClarty P A 2014 \textit{Rep. Prog. Phys.} \textbf{77} 056501.

\bibitem{5}
Starykh O A 2015 \textit{Rep. Prog. Phys.} \textbf{78} 052502.

\bibitem{6}
Collins M F and Petrenko O A 1997 \textit{Can. J. Phys.} \textbf{75} 605.

\bibitem{7}
Wiebe C R and Hallas A M 2015 \textit{APL Mater.} \textbf{3} 041519.

\bibitem{8}
Ramirez A P 1994 \textit{Annu. Rev. Mater. Sci.} \textbf{24} 453.

\bibitem{9}
Balents L 2010 \textit{Nature} \textbf{464} 199.

\bibitem{10}
Mendels P and Bert F 2016 \textit{C. R. Physique} \textbf{17} 455.

\bibitem{11}
Lee P A 2008 \textit{Science} \textbf{321} 1306.

\bibitem{12}
Zhou Y, Kanoda K and Ng T K 2017 \textit{Rev. Mod. Phys.} \textbf{89} 025003.

\bibitem{13}
Banerjee A, Yan J Q, Knolle J, Bridges C A, Stone M B, Lumsden M D, Mandrus D G, Tennant D A, Moessner R and Nagler S E 2017 \textit{Science} \textbf{356} 1055.

\bibitem{14}
Paddison J A M, Daum M, Dun Z L, Ehlers G, Liu Y H, Stone M B, Zhou H D and Mourigal M 2017 \textit{Nat. Phys.} \textbf{13} 117.

\bibitem{15}
Norman M R 2016 \textit{Rev. Mod. Phys.} \textbf{88} 041002.

\bibitem{16}
Mendels P and Bert F 2010 \textit{J. Phys. Soc. Jpn.} \textbf{79} 011001.

\bibitem{17}
Shores M P, Nytko E A, Barlett B M and Nocera D G 2005 \textit{J. Am. Chem. Soc.} \textbf{127} 13462.

\bibitem{18}
Krivovichev S V, Hawthorne F C and Williams P A 2017 \textit{Struct. Chem.} \textbf{28} 153.

\bibitem{19}
Zheng X G, Fijihala M, Kitajima S, Maki M,  Kato K, Takata M and Xu C N 2013 \textit{Phys. Rev. B} \textsc{87} 174102.

\bibitem{20}
Mori W and Yamaguchi K 1995 \textit{Mol. Cryst. Liq. Cryst}. \textbf{274} 113.

\bibitem{21}
Zheng X G and Otabe E S 2004 \textit{Solid State Commun}. \textbf{130} 107.

\bibitem{22}
Zheng X G, Hagihala M, Kawae T and Xu C N 2008 \textit{Phys. Rev. B} \textbf{77} 024418.

\bibitem{23}
Zheng X G, Hagihala M and Toriyi T 2007 \textit{J. Mag. Mag. Mater.} \textbf{310} 1288.

\bibitem{24}
Fujihala M, Hagihala M, Zheng X G and Kawae T 2010 \textit{Phys. Rev. B} \textbf{82} 024425.

\bibitem{25}
Hagihala M, Zheng X G, Kawae T and Sato T J 2010 \textit{Phys. Rev. B} \textbf{82} 214424.

\bibitem{26}
Zheng X G, Yamashita T, Hagihala M, Fujihala M and Kawae T 2009 \textit{ Physica B} \textbf{404} 680.

\bibitem{27}
Kotyuzhanskii B Y and Nikiforov D V 1991 \textit{J. Phys.: Condens. Matter} \textbf{3} 385.

\bibitem{28}
Yamazaki H 1995 \textit{J. Phys. Soc. Jpn.} \textbf{64} 2347.

\bibitem{29}
F\"{o}rster T, Garcia F A, Gruner T, Kaul E E, Schmidt B, Geibel C and Sichelschmidt J 2013 \textit{Phys. Rev. B} \textbf{87} 180401(R).

\bibitem{30}
Ishikawa H, Nakamura N, Yoshida M, Takigawa M, Babkevich P, Qureshi N, R{\o}nnow H M, Yajima T and Hiroi Z 2017 \textit{Phys. Rev. B} \textbf{95} 064408.

\bibitem{31}
Xiao F, Woodward F M, Landee C P, Turnbull M M, Mielke C, Harrison N, Lancaster T, Blundell S J, Baker P J, Babkevich P and Pratt F L 2009 \textit{Phys. Rev. B} \textbf{79} 134412.

\bibitem{33}
Gordon J E, Tan M L, Fisher R A and Phillips N E 1989 \textit{Soild State Commun.} \textbf{69} 625.

\bibitem{42}
Tari 2003 \textit{Specific Heat of Matter at Low Temperatures} (Imperial College Press).

\end{thebibliography}
\end{document}